\documentclass[showpacs, showkeys,twocolumn]{revtex4}
\begin{document}
\title{ Interplay between quark-antiquark and diquark
condensates in vacuum in a two-flavor Nambu-Jona-Lasinio
model\footnote{The project supported by the National Natural
Science Foundation of China under Grant No.10475113.} \\
}
\author{ZHOU Bang-Rong}
\affiliation{College of Physical Sciences,
Graduate School of the Chinese Academy of Sciences, Beijing
100049, China}
\affiliation{ CCAST (World Laboratory) P.O. Box
8730, Beijing 100080, China}
\date{}
\begin{abstract}
By means of a relativistic effective potential, we have
analytically researched competition between the quark-antiquark
condensates $\langle\bar{q}q\rangle$ and the diquark condensates
$\langle qq\rangle$ in vacuum in ground state of a two-flavor
Nambu-Jona-Lasinio (NJL) model and obtained the $G_S-H_S$ phase
diagram, where $G_S$ and $H_S$ are the respective four-fermion
coupling constants in scalar quark-antiquark channel and scalar
color anti-triplet diquark channel. The results show that, in the
chiral limit, there is only the pure $\langle\bar{q}q\rangle$
phase when $G_S/H_S>2/3$, and as $G_S/H_S$ decreases to
$2/3>G_S/H_S\geq 0$ one will first have a coexistence phase of the
condensates $\langle\bar{q}q\rangle$ and $\langle qq\rangle$ and
then a pure $\langle qq\rangle$ phase.  In non-zero bare quark
mass case, the critical value of $G_S/H_S$ at which the pure
$\langle\bar{q}q\rangle$ phase will transfer to the coexistence
phase of the condensates $\langle\bar{q}q\rangle$  and $\langle
qq\rangle$ will be less than 2/3. Our theoretical results,
combined with present phenomenological fact that there is no
diquark condensates in the vacuum of QCD, will also impose a real
restriction to any given two-flavor NJL model which is intended to
simulate QCD, i.e. in such model the resulting smallest ratio
$G_S/H_S$ after the Fierz transformations in the Hartree
approximation must be larger than 2/3. A few phenomenological
QCD-like NJL models are checked and analyzed.
\end{abstract}
 \maketitle
\section{Introduction\label{Intro}}
It is well known that in Quantum Chromodynamics (QCD) at
temperature $T=0$  and quark chemical potential $\mu=0$, i.e. in
vacuum, quarks are in chiral symmetry breaking phase owing to the
quark-antiquark condensates $\langle\bar{q}q\rangle\neq 0$
\cite{kn:1,kn:2,kn:3,kn:4,kn:5,kn:6,kn:7}.  Recently It has also
been clear that, at low temperature and high baryonic density, the
quarks will be in color superconducting phase owing to some
diquark condensates being non-zeroes \cite{kn:8,kn:9,kn:10,kn:11}.
Research on diquark condensates and color superconducting has also
been extended to  the region of low temperature and moderate quark
chemical potential where perturbative or present-day lattice
calculations are inaccessible, by means of different
phenomenological models including the Nambu-Jona-Lasinio
(NJL)-type models \cite{kn:2,kn:4,kn:12}. Since the interactions
between two quarks in the antitriplet of the color $SU_c(3)$ gauge
group are attractive, one can reasonably inquire if the diquark
condensates could be generated in vacuum. This problem has been
researched or touched on by means of some phenomenological models
\cite{kn:13,kn:14,kn:15}.  Some results from a few given NJL-type
models were obtained.  However, it seems that, for a thorough
understanding of the diquark condensates in a NJL model in vacuum,
especially of the interplay between the diquark and the
quark-antiquark condensates in vacuum, a general and systematic
analysis is still necessary. In addition, such research itself has
certainly theoretical interest if one treats the NJL model as a
useful field theory model. The NJL model as a phenomenological
description of QCD, is usually used in discussion of chiral
symmetry breaking but very little in the diquark condensation
problem in vacuum. Then what is the theoretical reason of this
situation? In fact, in any NJL model, owing to the Fierz
transformation \cite{kn:12}, one is always allowed to include both
the interactions of $(\bar{q}q)^2$-form and $(qq)^2$-form which
could lead to generation of the quark-antiquark and the diquark
condensates respectively. As long as applying a NJL model to
diquark condensation problem in vacuum, one will inevitably face
to mutual competition between the diquark condensates and the
quark-antiquark condensates.  A serious research on such mutual
competition in a NJL model will lead to some condition in which
the diquark condensates could be generated, or say, removed in
vacuum. By this condition, combined with the phenomenological
requirements of QCD in vacuum, one will be able to answer why in
conventional NJL model description of QCD in vacuum one can
completely neglect existence of the diquark condensates. In fact,
the condition to remove the diquark condensates in vacuum will
become a useful restriction to any "realistic" NJL model.\\
\indent Motivated by the above ideas, in this paper we will make a
general analysis of the mutual competition between the
quark-antiquark and diquark condensates in a NJL-type model in
vacuum. The paper is arranged as follows. In Sect.\ref{model} we
will present the model and derive its effective potential, in
Sect.\ref{ground} determine the ground states in different
conditions and in Sect.\ref{concl} come to our conclusions and
discussions.
\section{Model and Effective potential\label{model}}
we start with a two-flavor NJL model which could be used to
simulate QCD and derive its effective potential containing the
order parameters respectively corresponding to quark-antiquark and
diquark condensates in  mean field approximation. As is well
known, in the mean field approximation, there is so called Fierz
ambiguity \cite{kn:11,kn:16}. However, throughout the whole
discussions in this paper, we will follow Ref.\cite{kn:12} and use
the Fierz-transformed four-fermion couplings in the Hartree
approximation to avoid double counting.  To simplify the problem
and to make it possible to conduct the demonstration in a
completely analytic way, we will first omit the small bare masses
of the $u$ and $d$ quarks (i.e. take the chiral limit). Thus the
Lagrangian of the two-flavor NJL model may be written in the
well-known form by
\begin{eqnarray}
{\cal L}&=&\bar{q}i\not\!\partial q
+G_S[(\bar{q}q)^2+(\bar{q}i\gamma_5\vec{\tau}q)^2]\nonumber\\
&&+H_S\sum_{A=2,5,7}(\bar{q}i\gamma_5\tau_2\lambda_Aq^C)
    ({\bar{q}}^Ci\gamma_5\tau_2\lambda_Aq),
\end{eqnarray}%%(1)%%
with the quark fields $q$ in the $SU_f(2)$ doublet and the
$SU_c(3)$ triplets, i.e.
$$q=\left(
\begin{array}{c}
  u_i\\
  d_i \\
\end{array}
\right)\;\; i=r,g,b,
$$
where the subscripts $i=r, g, b$ denote the three colors (red,
green and blue) of quarks. In Eq.(1) we have used the denotations
$\not\!{\partial}\equiv \gamma^{\mu}\partial_{\mu}$, Pauli
matrices $\vec{\tau}=(\tau_1,\tau_2,\tau_3)$ acting in the
two-flavor space, the Gell-Mann matrices $\lambda_A$ acting in the
three-color space and the charge conjugations $q^C$ and
$\bar{q}^C$ of the quark fields. The Lagrangian (1) is
$SU_c(3)\otimes SU_{fL}(2)\otimes SU_{fR}(2)\otimes
U_f(1)$-invariant. For the sake of making a general discussion of
mutual competition between the diquark and the quark-antiquark
condensates in the model, we temporarily view the positive
coupling constants $G_S$ and $H_S$ as independent and changeable
parameters. Assume that the four-fermion interactions can lead to
the scalar quark-antiquark condensates
\begin{equation}
    \langle\bar{q}q\rangle=\phi
\end{equation}%%(2)%%
which will break the chiral symmetries $SU_{fL}(2)\otimes
SU_{fR}(2)$ down to $SU_{fV}(2)$ and induce three massless pions
as corresponding Goldstone bosons. We also suppose that the
four-fermion interactions could generate the scalar
color-antitriplet diquark and color-triplet di-antiquark
condensates (after a global $SU_c(3)$ transformation)
\begin{equation}
\langle{\bar{q}}^Ci\gamma_5\tau_2\lambda_2q\rangle=\delta,\;
\langle\bar{q}i\gamma_5\tau_2\lambda_2q^C\rangle=\delta^*.
\end{equation}%%(3)%%
It is indicated that in view of the structure of the matrix
$\lambda_2$ in $\delta$ and $\delta^*$ in Eq.(3), only the red and
green quarks participate in the diquark condensates and the blue
quarks do not enter them \cite{kn:12}. For acquiring a qualitative
understanding of the diquark condensates in vacuum, we can assume,
similar to the BCS ansatz in superconducting theory, physical
vacuum as a coherent state of red and green up and down quarks
with zero total momentum \cite{kn:9} which contains both
quark-quark paring and antiquark-antiquark paring. It is indicated
that the NJL model in vacuum i.e. at $T=\mu=0$ is a relativistic
quantum field theory. Hence, despite of absence of net quarks in
vacuum, in such theory it is possible that the quark-quark
condensates and the antiquark-antiquark condensates are generated
simultaneously in vacuum, such as shown by $\delta$ and $\delta^*$
in Eq.(3). This case is also similar to generation of
quark-antiquark condensates in vacuum, where physical vacuum can
be assumed to be a coherent state composed of quark-antiquark
pairs with zero total momentum\cite{kn:2,kn:9}. As in the
two-flavor color superconducting theory, the condensates (3) will
leave the chiral symmetries $SU_{fL}(2)\otimes SU_{fR}(2)$, a
"rotated" electric charge $U_{\tilde{Q}}(1)$ and a "rotated" quark
number $U'_q(1)$ symmetry unbroken, but will break the $SU_c(3)$
symmetry down to a $SU_c(2)$ symmetry \cite{kn:12}. As a result,
if the $SU_c(3)$ group is gauged, then five of the eight gluons
will receive a mass through the Higgs mechanism, and if the
$SU_c(3)$ group is global, then five massless diquark excitations,
which could be some combinations of quark-quark pairs and
antiquark-antiquark pairs, will arise as the Goldstone bosons of
spontaneous breaking of $SU_c(3)$. We note that the flavor chiral
symmetry breaking induced by the quark-antiquark condensates and
the color symmetry breaking induced by the diquark condensates are
independent each other, so even if the two condensates coexist in
ground state, the total physical effects will be only a simple
adding up of the effects from the two-forms of symmetry breaking.
However, in this paper, we will omit further discussions of the
possible physical consequences of the diquark condensates in
vacuum, and pay our main attention to the conditions to generate
the two-forms of condensates and mutual competition between them.
Now define that
\begin{equation}
\sigma=-2G_S\phi, \;\;
\Delta=-2H_S\delta,\;\;\Delta^*=-2H_S\delta^*
\end{equation}%%(4)%%
then in mean-field (Hartree) approximation \cite{kn:12, kn:17} and
in the Nambu-Gorkov (NG) basis \cite{kn:18} with the denotations
$$\Psi=\frac{1}{\sqrt{2}}\left(
\begin{array}{c}
  q \\
  q^C \\
\end{array}
\right),\;\; \bar{\Psi}=\frac{1}{\sqrt{2}}\left(%
\begin{array}{cc}
  \bar{q}, & \bar{q}^C \\
\end{array}
\right),
$$
the Lagrangian (1) may be rewritten by
\begin{equation}
{\cal L} =
   \bar{\Psi}S^{-1}(x)\Psi-\sigma^2/4G_S-|\Delta|^2/4H_S.
\end{equation}%%(5)%%
In momentum space, the inverse propagator $S^{-1}(x)$ has the
expression
\begin{equation}
S^{-1}(p)=\left(
\begin{array}{cc}
  \not\!{p}-\sigma & -i\gamma_5\tau_2\lambda_2\Delta \\
  -i\gamma_5\tau_2\lambda_2\Delta^* & \not\!{p}-\sigma \\
\end{array}
\right).
\end{equation}%%(6)%%
Different from the thermodynamic potential at finite temperature
and finite chemical potential case, the effective potential in
vacuum corresponding to the Lagrangian (5) will be
relativistic-invariant. It can be obtained by a direct
generalization of the conventional one order-parameter effective
potential formula \cite{kn:7,kn:19} to the case with two
order-parameters and in the NG basis. It can be expressed by
\begin{equation}
V(\sigma,\Delta)=\frac{\sigma^2}{4G_S}+\frac{|\Delta|^2}{4H_S}+i
\int\frac{d^4p}{(2\pi)^4}\frac{1}{2}\textrm{Tr}\ln S^{-1}(p)S_0(p)
\end{equation}%%(7)%%
where $S_0(p)$ represents the propagator for massless quarks in
the NG basis and the Tr is to be taken over flavor, color,
Nambu-Gorkov and Dirac spin degrees of freedom. The factor $1/2$
is due to use of the NG basis. Since the blue quarks do not enter
the diquark condensates, we will have
\begin{eqnarray}
&&\mathrm{Tr}\ln S^{-1}(p)S_0(p)=\ln
\mathrm{Det}S^{-1}(p)S_0(p)\nonumber\\
&&=\ln\mathrm{Det}S^{-1}(p)S_0(p)|_{(r,g)}+
\ln\mathrm{Det}S^{-1}(p)S_0(p)|_b \nonumber\\
&&=\mathrm{Tr}\ln S^{-1}(p)S_0(p)|_{(r,g)}+\mathrm{Tr}\ln
S^{-1}(p)S_0(p)|_b,
\end{eqnarray}%%(8)%%
where subscripts $(r,g)$ and $b$ mean that the corresponding
matrices are limited to (red, green) and blue degrees of freedom
in color space respectively.  By using the mathematical formula
$$\textrm{Det}\left(
\begin{array}{cc}
  A & B \\
  C & D \\
\end{array}
\right)=\mathrm{Det}[-CB+CAC^{-1}D],$$ where in the above
$2\times2$ matrix, $A$, $B$, $C$ and $D$ can be operators
(matrices), we obtain
\begin{widetext}
\begin{eqnarray}
\textrm{Tr}\ln S^{-1}(p)S_0(p)|_{(r,g)}&=&\ln
\mathrm{Det}\frac{\not\!p}{p^2+i\varepsilon}\left(
\begin{array}{cc}
  \not\!{p}-\sigma & -i\gamma_5\tau_2\tau^c_2\Delta \\
  -i\gamma_5\tau_2\tau^c_2\Delta^* & \not\!{p}-\sigma \\
\end{array}
\right)=16\ln
\frac{p^2-\sigma^2-|\Delta|^2+i\varepsilon}{p^2+i\varepsilon},\\
%%(9)%%
\textrm{Tr}\ln S^{-1}(p)S_0(p)|_b&=&\ln
\mathrm{Det}\frac{\not\!p}{p^2+i\varepsilon}\left(
\begin{array}{cc}
  \not\!{p}-\sigma & 0 \\
  0 & \not\!{p}-\sigma \\
\end{array}
\right)\otimes(\mathbf{1}^f_2)=8\ln\frac{p^2-\sigma^2+i\varepsilon}{p^2+i\varepsilon},
\end{eqnarray}%%(10)%%
where $\tau^c_2=\tau_2$ but it is now acting in the $(r,g)$
two-color space and $(\mathbf{1}^f_2)$ represents the $2\times2$
unit matrix in the two-flavor space. Substituting Eqs.(9) and (10)
into Eq.(8), we may transform Eq.(7) to
\begin{equation}
V(\sigma, \Delta)=\frac{\sigma^2}{4G_S}+\frac{|\Delta|^2}{4H_S}+8i
\int\frac{d^4p}{(2\pi)^4}\ln
\frac{p^2-\sigma^2-|\Delta|^2+i\varepsilon}{p^2+i\varepsilon}
+4i\int\frac{d^4p}{(2\pi)^4}\ln
\frac{p^2-\sigma^2+i\varepsilon}{p^2+i\varepsilon}
\end{equation}%%(11)%%
By using the formula
$$I(a^2)=i\int\frac{d^4p}{(2\pi)^4}\ln\frac{p^2-a^2+i\varepsilon}{p^2+i\varepsilon}
=\int_0^{a^2}d u^2\frac{d I(u^2)}{du^2}
$$
it is easy to complete the integrations and obtain the final
explicit expression of the effective potential
\begin{equation} V(\sigma,
|\Delta|)=\frac{\sigma^2}{4G_S}+\frac{|\Delta|^2}{4H_S}-\frac{1}{4\pi^2}
\left[(3\sigma^2+2|\Delta|^2)\Lambda^2-
\frac{\sigma^4}{2}\left(\ln\frac{\Lambda^2}{\sigma^2}+\frac{1}{2}\right)-
(\sigma^2+|\Delta|^2)^2\left(\ln\frac{\Lambda^2}{\sigma^2+|\Delta|^2}+\frac{1}{2}\right)\right],
\end{equation}%%(12)%%
\end{widetext}
where we have made the Wick rotation of the integration variable
$p^0$ and introduced the 4D Euclidean squared momentum cutoff
$\Lambda^2$ which is assumed to satisfy the conditions
$\Lambda^2\gg \sigma^2$ and $\Lambda^2\gg \sigma^2+|\Delta|^2$. It
should be indicated that because the 4D Euclidean squared momentum
cutoff is used, the effective potential expressed by Eq.(12) will
be Lorentz-invariant and this feature is essential for description
of the diquark condensates in vacuum. As a comparison, we note
that a similar expression to Eq.(12) can be obtained by taking the
limit $T=\mu=0$ in some corresponding thermodynamical potential
but a 3D momentum cutoff is usually used there. Such a limit
expression of the thermodynamical potential, rigorously speaking,
is not suitable to the problem of the diquark condensates in
vacuum. Since the 3D momentum cutoff breaks the Lorentz
invariance, the relevant discussions will be theoretically
inconsistent to generating mechanism of the diquark condensates in
vacuum which must be built on a relativistic basis, as stated
above. Hence only the effective potential (12) with the 4D
momentum cutoff will give a rigorous and consistent relativistic
description of the problem involving the diquark condensates in
vacuum.

\section{Ground states\label{ground}}
\indent The relativistic effective potential given by Eq.(12)
contains two order parameters $\sigma$ and $|\Delta|$.  A obvious
advantage of the expression (12) is that one could analytically
find out the minimums of $V(\sigma, |\Delta|)$ and then determine
the ground state of the system. Simultaneously, one will be able
to examine the mutual competition between $\sigma$ and $|\Delta|$
when the values of the coupling constants $G_S$ and $H_S$ are
changed. The extreme value conditions $\partial V(\sigma,
|\Delta|)/\partial\sigma=0$ and $\partial V(\sigma,
|\Delta|)/\partial|\Delta|=0$ will separately lead to the
equations
\\
\begin{equation}
\sigma\left(\frac{1}{2G_S}-\frac{3\Lambda^2}{2\pi^2}+
\frac{\sigma^2}{2\pi^2}\ln\frac{\Lambda^2}{\sigma^2}+
\frac{\sigma^2+|\Delta|^2}{\pi^2}\ln\frac{\Lambda^2}{\sigma^2+|\Delta|^2}\right)
=0
\end{equation}%%(13)%%
and
\begin{equation}
|\Delta|\left(\frac{1}{2H_S}-\frac{\Lambda^2}{\pi^2}+
\frac{\sigma^2+|\Delta|^2}{\pi^2}\ln\frac{\Lambda^2}{\sigma^2+|\Delta|^2}\right)=0.
\end{equation}%%(14)%%
It is easy to see that Eqs.(13) and (14) may have nonzero $\sigma$
and $|\Delta|$ solutions if
\begin{equation}
G_S\Lambda^2>\pi^2/3
\end{equation}%%(15)%%
and
\begin{equation}
H_S\Lambda^2>\pi^2/2.
\end{equation}%%(16)%%
The two conditions could be satisfied when either the momentum
cutoff $\Lambda$ is large enough for given coupling constants
$G_S$ and $H_S$ or $G_S$ and $H_S$ are large enough for a given
$\Lambda$. From the second order derivatives of
$V(\sigma,|\Delta|)$ denoted by
\begin{widetext}
\begin{eqnarray}
A&\equiv&\frac{\partial^2V}{\partial\sigma^2}=\frac{1}{2G_S}-\frac{3\Lambda^2}{2\pi^2}+
\frac{3\sigma^2}{2\pi^2}\left(\ln\frac{\Lambda^2}{\sigma^2}-2\right)
+\frac{3\sigma^2+|\Delta|^2}{\pi^2}\ln\frac{\Lambda^2}{\sigma^2+|\Delta|^2},\nonumber\\
B&\equiv&\frac{\partial^2V}{\partial\sigma\partial|\Delta|}=
\frac{\partial^2V}{\partial|\Delta|\partial\sigma}=
2\frac{\sigma|\Delta|}{\pi^2}\left(\ln\frac{\Lambda^2}{\sigma^2+|\Delta|^2}-1\right),\nonumber\\
C&\equiv&\frac{\partial^2V}{\partial|\Delta|^2}=
\frac{1}{2H_S}-\frac{\Lambda^2+2|\Delta|^2}{\pi^2}+
\frac{\sigma^2+3|\Delta|^2}{\pi^2}\ln\frac{\Lambda^2}{\sigma^2+|\Delta|^2},
\end{eqnarray}%%(17)%%
\end{widetext}
we may define the expression
\begin{equation}
K\equiv\left|\begin{array}{cc}
  A & B \\
  B & C \\
\end{array}\right|=AC-B^2.
\end{equation}%%(18)%%
The equations (13) and (14) have four different solutions which
will be discussed successively as follows.\\
1) $(\sigma,|\Delta|)=(0,0)$. It is a maximum point of
$V(\sigma,|\Delta|)$, since in this case we have
$$A=1/2G_S-3\Lambda^2/2\pi^2<0,$$
$$K=AC=A(1/2H_S-\Lambda^2/\pi^2)>0,$$ if considering the necessary
conditions (15) and (16) for existence of non-zero $\sigma$ and $|\Delta|$ solutions.\\
2) $(\sigma,|\Delta|)=(\sigma_1,0)$, where $\sigma_1\neq 0$ and
obeys the equation
\begin{equation}
1/2G_S-3\Lambda^2/2\pi^2+
3\sigma_1^2\ln(\Lambda^2/\sigma_1^2)/2\pi^2=0.
\end{equation}%%(19)%%
It may be obtained by means of Eq.(19) that when $\Lambda$ is
large, we have
$$A=3\sigma_1^2[\ln(\Lambda^2/\sigma_1^2)-1]/\pi^2>0$$ and
$$
K=A\left(\frac{1}{2H_S}-\frac{1}{3G_S}\right)\left\{\begin{array}{ccc}
  >0 & \mbox{if} & G_S/H_S>2/3 \\
  \leq 0 & \mbox{if} & G_S/H_S\leq 2/3.\\
\end{array}\right.
$$
They indicate that $(\sigma_1,0)$ will be a minimum point of
$V(\sigma,|\Delta|)$ when $G_S/H_S>2/3$, however, it will not be
an extreme value point of $V(\sigma,|\Delta|)$ when
$G_S/H_S\leq2/3$.\\
3) $(\sigma,|\Delta|)=(0,\Delta_1)$, where $\Delta_1\neq 0$ and
obeys the equation
\begin{equation}
1/2H_S-\Lambda^2/\pi^2+
\Delta_1^2\ln(\Lambda^2/\Delta_1^2)/\pi^2=0.
\end{equation}%%(20)%%
In this case we obtain that
$$A=1/2G_S-3/4H_S-
(\Lambda^2/\pi^2-1/2H_S)/2,$$
$$
K=A\cdot 2\Delta_1^2[\ln(\Lambda^2/\Delta_1^2)-1]/\pi^2.
$$
Obviously, the sign of $K$ depends on $A$'s one. We will have
$A>0$, if
\begin{equation}
G_S/H_S<1/(1+H_S\Lambda^2/\pi^2).
\end{equation}%%(21)%%
We note that
\begin{equation} 1/(1+H_S\Lambda^2/\pi^2)<2/3,
\end{equation}%%(22)%%
owing to Eq.(16). Equation(21) certainly includes the case of
$G_S=0$. Thus $(\sigma,|\Delta|)=(0,\Delta_1)$ will be a minimum
point of $V(\sigma,|\Delta|)$ only if Eq.(21) is satisfied, i.e.
either the scalar quark-antiquark interactions do not exist or
they are very weak, or in other words, the scalar diquark
interactions are strong enough opposite to the scalar
quark-antiquark interactions. \\
4)$(\sigma,|\Delta|)=(\sigma_2,\Delta_2)$.  The none-zero
$\sigma_2$ and $\Delta_2$ obey the equations
\begin{equation}
\frac{1}{2G_S}-\frac{3\Lambda^2}{2\pi^2}+
\frac{\sigma_2^2}{2\pi^2}\ln\frac{\Lambda^2}{\sigma_2^2}+
\frac{\sigma_2^2+\Delta_2^2}{\pi^2}\ln\frac{\Lambda^2}{\sigma_2^2+\Delta_2^2}=0,
\end{equation}%%(23)%%
\begin{equation}
\frac{1}{2H_S}-\frac{\Lambda^2}{\pi^2}+
\frac{\sigma_2^2+\Delta_2^2}{\pi^2}\ln\frac{\Lambda^2}{\sigma_2^2+\Delta_2^2}=0.
\end{equation}%%(24)%%
Now from the calculated results by using Eqs.(23) and (24) that
$$A=\frac{\sigma^2_2}{\pi^2}\left[3\left(\ln\frac{\Lambda^2}{\sigma_2^2+\Delta_2^2}-1\right)
+\ln\frac{\sigma_2^2+\Delta_2^2}{\sigma^2_2}\right]>0,
$$
$$K=\frac{2\sigma_2^2\Delta_2^2}{\pi^4}\left(\ln\frac{\Lambda^2}{\sigma_2^2+\Delta_2^2}-1\right)
\left(\ln\frac{\Lambda^2}{\sigma_2^2}-1\right)>0,
$$
we may deduce that so long as a non-zero solution
$(\sigma_2,\Delta_2)$ exists, it will be a minimum point of
$V(\sigma,|\Delta|)$. The remaining problem is to find out the
condition in which $\sigma_2$ and $\Delta_2$ obeying Eqs.(23) and
(24) could simultaneously be equal to non-zero. For this purpose,
we note the fact that the function
$$f(a^2)=a^2\ln\frac{\Lambda^2}{a^2}$$
has the feature that
$$\frac{df(a^2)}{da^2}=\ln\frac{\Lambda^2}{a^2}-1>0,\;\;
\mathrm{if}\;\; \frac{\Lambda^2}{a^2}>e,
$$
i.e. $f(a^2)$ will be a monotonically increasing function of $a^2$
for large enough $\Lambda^2$. Hence, if
$\Lambda^2/\sigma_2^2>\Lambda^2/(\sigma_2^2+\Delta_2^2)>e$, we
will have the inequality
\begin{equation}
\sigma_2^2\ln(\Lambda^2/\sigma_2^2)<
(\sigma_2^2+\Delta_2^2)\ln[\Lambda^2/(\sigma_2^2+\Delta_2^2)].
\end{equation}%%(25)%%
Applying Eq.(25) to Eqs.(23) and (24), we are led to the condition
\begin{equation}
G_S/H_S<2/3.
\end{equation}%%(26)%%
On the other hand, by Eqs.(23) and (24) we have the inequality
$$1/2G_S-3\Lambda^2/2\pi^2<1/2H_S-\Lambda^2/\pi^2$$
which may be changed into
\begin{equation}
G_S/H_S>1/(1+H_S\Lambda^2/\pi^2).
\end{equation}%%(27)%%
Hence only if Eqs.(26) and (27) are satisfied, we can have a
non-zero solution $(\sigma_2, \Delta_2)$ as a minimum point of
$V(\sigma,|\Delta|)$. It may be seen by comparing Eq.(21) with
Eq.(27) that the minimum points $(0,\Delta_1)$ and $(\sigma_2,
\Delta_2)$ can not coexist, so in each case the only minimum point
(for $\sigma >0$) will correspond to the ground state of the
system, especially the solution $(\sigma_2, \Delta_2)$ implies
coexistence of the
quark-antiquark and diquark condensates in the ground state.\\
\indent All the above results can be summarized as follows.
Locations of the minimum points of the effective potential
$V(\sigma,|\Delta|)$ depend on relative values of $G_S$ and $H_S$.
The minimum points will respectively be at
\begin{widetext}
\begin{equation}
(\sigma,|\Delta|)=\left\{\begin{array}{lc}
  (0, & \Delta_1) \\
  (\sigma_2, & \Delta_2) \\
  (\sigma_1, & 0) \\
\end{array}\right.\;\;\mbox{if}\;\;
\left\{\begin{array}{cccl}
  &0&\leq&G_S/H_S<1/(1+H_S\Lambda^2/\pi^2) \\
  &1/(1+H_S\Lambda^2/\pi^2)&<&G_S/H_S<2/3 \\
  &&&G_S/H_S>2/3 \\
\end{array}\right.,
\end{equation}%%(28)%%
\end{widetext}
where $\sigma_1$ and $\Delta_1$ are determined by Eqs.(19) and
(20) separately, and $(\sigma_2,\Delta_2)$ by Eqs.(23) and (24).
Combining Eq.(28) with Eqs.(15) and (16), we have shown in Fig.1
the corresponding regions to the three different phases in the
coupling constants $G_S-H_S$ plane.  \\
\vspace{3cm} \begin{center}(location of
Fig.1)\end{center}\vspace{3cm} It may be seen that in the region
with non-zero condensates, the pure quark-antiquark condensate
phase and the pure diquark condensate phase is separated by a
coexistence phase of the two condensates.  in each phase, either
($0$,$\Delta_1$) or ($\sigma_2$,$\Delta_2$) or ($\sigma_1$,0) is
the only minimum point of $V(\sigma,|\Delta|)$ and no other
minimum point arises, so there is no metastable state in the model.
We indicate that the above $G_S-H_S$ phase diagram in the vacuum of
a NJL model is first obtained.\\
\indent The results show that the pure quark-antiquark condensates
phase and the phase with diquark condensates is distinguished
merely by the ratio $G_S/H_S=2/3$ and the diquark condensates
could emerge only if $H_S\Lambda^2/\pi^2>1/2$ and $G_S/H_S$ is
below the critical value 2/3. From the expressions (11) and (12)
of the effective potential $V(\sigma,|\Delta|)$ and the follow-up
discussions it is not difficult to see that the critical value 2/3
reflects the following fact: in the two-flavor NJL model only the
two colors (red and green) of quarks anticipate in the diquark
condensates but all the three colors (red, green and blue) of
quarks get into the quark-antiquark condensates.
\section{Conclusions and discussions\label{concl}}
 We have analyzed the effective
potential of a two-flavor NJL model at $T=\mu=0$ which contains
two order parameters $\sigma$ and $|\Delta|$ separately
corresponding to quark-antiquark and diquark condensates and
proven that the mutual competition between $\sigma$ and $|\Delta|$
is decided by the ratio of the scalar quark-antiquark and the
scalar diquark four-fermion coupling constant $G_S$ and $H_S$. The
results indicate that even in vacuum, the diquark condensates
could either exist alone or coexist with quark-antiquark
condensates when $G_S/H_S<2/3$ and $H_S$ is large enough. However,
only the pure quark-antiquark condensates could be generated when
$G_S/H_S>2/3$. The last conclusion coincides with the result
obtained in Ref. \cite{kn:14} based on a random matrix model,
where the critical value of $G_S/H_S$ above which no diquark
condensates arise is $2/N_c$ for a color $SU_c(N_c)$ group.  In
fact, generalization of the analysis in present paper to $SU_c(N)$
case is direct. In doing so, not only the critical value $2/N_c$
will be derived but also a complete $G_S-H_S$ phase diagram
similar to Fig.1 can be obtained. The conclusion that a very large
coupling constant $H_S$ in an NJL model  could lead to diquark
condensates in vacuum is also touched on in Ref.\cite{kn:15}, but
no further details were given there. It is also indicated that, in
the above both models no corresponding $G_S-H_S$ phase diagram was given.\\
\indent The above conclusions are reached in the chiral limit. If
a non-zero degenerate bare quark mass $m_u=m_d=m$ is included,
then the effective potential will have a little more complicated
expression thus a complete analytic demonstration will become more
difficult. However, the essential part of the discussions
including determination of the conditions of emergence of the
minimum points $(\sigma_1,0)$ and $(\sigma_2,\Delta_2)$ can still
be conducted analytically and the relevant essential conclusions
reached in the $m=0$ case will keep unchanged. In particular, we
can definitely give the critical value of $G_S/H_S$ at which the
pure quark-antiquark condensate phase will transfer to the
coexistence phase of quark-antiquark and diquark condensates, but
now the value will change from 2/3 into
$2\sigma_{min}/3(\sigma_{min}+m)$, where $\sigma_{min}$ is the
value of the order-parameter related to the quark-antiquark
condensates in the ground state. Because the extra factor
$\sigma_{min}/(\sigma_{min}+m)\leq 1$, so in nor-zero bare quark
mass case, appearance of the diquark condensates requires
stronger diquark interactions than the ones in the chiral limit. \\
\indent In our discussions, the coupling constants $G_S$ and $H_S$
have been viewing as independent and changeable parameters, this
is only a theoretical assumption for the purpose of exclusively
researching mutual competition between the diquark and the
quark-antiquark condensates in the model. In fact, $G_S$ and $H_S$
are interrelated via the Fierz transformations.  Therefore, for a
given NJL model in advance, the ratio $G_S/H_S$ in Eq.(28) must be
understood as its derived value in the Hartree approximation after
the Fierz transformations are implemented.\\
\indent By the above results, it seems that, theoretically, a
general two-flavor  NJL model does not remove  existence of
diquark condensates in vacuum, as long as the diquark interactions
are strong enough. However, this can not be realistic case. Since
a NJL model is usually used as a low energy phenomenological
description of QCD, its coupling forms and strengths must be
restricted by the underlying QCD theory and/or phenomenology,
hence the ratio $G_S/H_S$ can not be arbitrary. If the underlying
Lagrangian or all the couplings of $(\bar{q}q)^2$-form are known,
then via the Fierz transformations, the effective coupling $G_S$
including direct and exchange interactions and the coupling $H_S$
in the diquark channel can be fixed uniquely in the Hartree
approximation, and so is the ratio $G_S/H_S$. However, when the
underlying Lagrangian is unknown, one can write down only the
partial coupling terms of $(\bar{q}q)^2$-form by symmetries and/or
phenomenology, then the effective $G_S$ and $H_S$ derived by the
Fierz transformations (still in the Hartree approximation) will
not be unique, since they will change with different selections of
the undetermined coupling terms of $(\bar{q}q)^2$-form. \\
\indent However, our theoretical results (28), combined with
phenomenological fact that there is no diquark condensates in the
vacuum of QCD, will be able to place a useful restriction to the
above different selections. In fact, for any NJL model which is
intended to simulate QCD, whatever initial coupling terms as a
starting point of the model building are selected, via the Fierz
transformations, the derived least value of the effective ratio
$G_S/H_S$ of the positive scalar quark-antiquark and the positive
scalar diquark coupling constants in the Hartree approximation
must be larger than 2/3. This restriction is applicable to any
realistic NJL model. \\
\indent Two examples of the models which obey the above
restriction are the models where the four-fermion interactions are
assumedly induced by heavy gluon exchange or by instantons in QCD.
These interactions can considered as some underlying more
microscopic ones. In both models, the resulting ratio $G_S/H_S$
via the Fierz transformations are equal to 4/3 \cite{kn:12} which
is obviously larger than the critical value 2/3 for generation of
the diquark condensates. So in these models we will have only the
quark-antiquark condensates surviving. In Ref.\cite{kn:13} a
slightly different conclusion was reached in an instanton-induced
model with fixed values of $G_S$ and $H_S$. There for $N_c=3$
case, in the ground state similarly only the quark-antiquark
condensates exist, but a pure diquark condensates also exist in a
metastable state. However, it is noted that in that model the
corresponding ratio $G_S/H_S$ was not taken to be the conventional
value 4/3 and a further relation between
the coupling constants and the instanton density was used.\\
\indent Another more phenomenological model is the conventional
chiral $SU_{fL}(2)\otimes SU_{fR}(2)$-invariant four-fermion model
with interactions \cite{kn:4}
$${\cal L}_{int}=g[(\bar{q}q)^2+(\bar{q}i\gamma_5\vec{\tau}q)^2].
$$
Via the Fierz transformations we may obtain the effective
Lagrangian
$${\cal L}_{int}^{eff}={\cal L}_{q\bar{q}}+{\cal L}_{qq}
$$
with
\begin{eqnarray*}
\mathcal{L}_{q\bar{q}}&=&
G_S(\bar{q}q)^2+G_P(\bar{q}i\gamma_5q)^2+
G_P^\tau(\bar{q}i\gamma_5\vec{\tau}q)^2+\cdots, \\
\mathcal{L}_{qq}&=&H_S(\bar{q}i\gamma_5\tau_2\lambda_{A}q^C)
    ({\bar{q}}^Ci\gamma_5\tau_2\lambda_{A}q)\\
    &&+H_P(\bar{q}\tau_2\lambda_{A}q^C)
    ({\bar{q}}^C\tau_2\lambda_{A}q)+\cdots,
\end{eqnarray*}
where the ellipses indicate the possible vector, axial-vector and
tensor coupling terms.  It is emphasized that the effective
Lagrangian ${\cal L}_{int}^{eff}$ is used only in the Hartree
approximation \cite{kn:12}. For the two-flavor and three-color
case, we get that the corresponding effective coupling constants
$$G_S=G_P^\tau=11g/12,\;\; G_P=g/12,\;\;
H_S=-H_P=g/4 $$ Obviously, owing to that $G_S/H_S=11/3>2/3$, it is
impossible to generate the scalar diquark condensates in this
model. The maximal attractive channels are still the original
starting point, i.e. the coupling terms $(\bar{q}q)^2$ and
($\bar{q}i\gamma_5\vec{\tau}q)^2$. The possible condensates will
be $\langle\bar{q}q\rangle$ (and/or the $\pi$ condensates
$\langle\bar{q}i\gamma_5\tau_aq\rangle$) and
they are related to spontaneous chiral symmetry breaking only.\\
\indent In brief, in a few known possibly realistic QCD-analogous
two-flavor NJL models, owing to that the condition to generate the
diquark condensates in vacuum is not satisfied, there could be
only the quark-antiquark condensates in ground states of these
models at $T=\mu=0$.  These models are merely suitable for
description of chiral symmetry breaking and have nothing to do
with color superconductivity at $T=\mu=0$. This certainly reflects
the reality of QCD in vacuum.\\

\begin{center}FIGURE\end{center}
Fig.1  The corresponding regions to the three different phases
($0$, $\Delta_1$), ($\sigma_2$, $\Delta_2$) and ($\sigma_1$,0) in
$G_S-H_S$ plane (in dimensionless couplings $y=G_S\Lambda^2/\pi^2$
and $x=H_S\Lambda^2/\pi^2$).
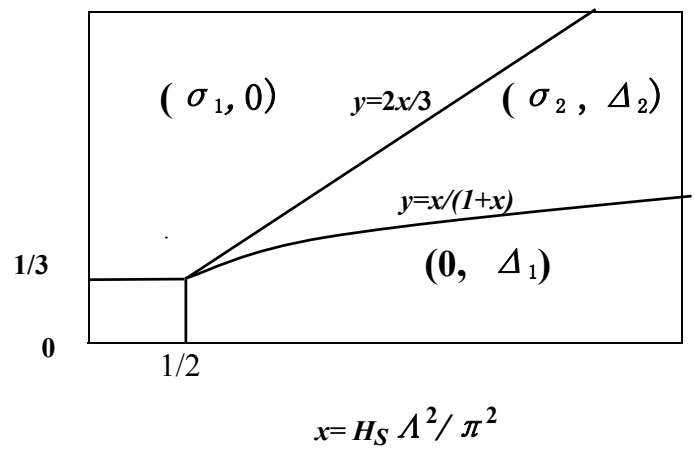
\end{document}